\documentclass[reprint,
 superscriptaddress,
 groupedaddress,
preprintnumbers,
nofootinbib,
floatfix,
longbibliography,
 amsmath,amssymb,
 aps,
 showpacs,
 prd,
floatfix,
]{revtex4-2}

\usepackage{color, colortbl}
\definecolor{Gray}{gray}{0.9}
\usepackage{siunitx}
\usepackage{makecell}
\usepackage{multirow}

\usepackage{booktabs} 
\usepackage{hyperref}
\hypersetup{
  colorlinks   = true,    
  urlcolor     = blue,    
  linkcolor    = blue,    
  citecolor    = red      
}
\usepackage{graphicx}
\usepackage{dcolumn}
\usepackage{bm}

\usepackage{changes}

\def \Nsh{N_{\rm shift}}
\def \Tsh{T_{\rm shift}}

\usepackage{xcolor}

\begin{document}

\preprint{APS/123-QED}
\title{Obtaining Statistical Significance of Gravitational Wave Signals in Hierarchical Search}

\author{Kanchan Soni}
\email{kanchansoni@iucaa.in}
\affiliation{
Inter-University Centre for Astronomy and Astrophysics, Pune 411007, India
}%

\author{Sanjeev Dhurandhar}
\email{sanjeev@iucaa.in}
\affiliation{
Inter-University Centre for Astronomy and Astrophysics, Pune 411007, India
}%

\author{Sanjit Mitra}
\email{sanjit@iucaa.in}
\affiliation{
Inter-University Centre for Astronomy and Astrophysics, Pune 411007, India
}%

\begin{abstract}
Gravitational Wave (GW) astronomy has experienced remarkable growth in recent years, driven by advancements in ground-based detectors. While detecting compact binary coalescences (CBCs) has become routine, searching for more complex ones, such as mergers involving eccentric and precessing binaries and sub-solar mass binaries, has presented persistent challenges. These challenges arise from using the standard matched filtering algorithm, whose computational cost increases with the dimensionality and size of the template bank. This urges the pressing need for faster search pipelines to efficiently identify GW signals, leading to the emergence of the hierarchical search strategy that reduces the computational cost of matched filtering in the search. This method looks for potential candidate events using a coarse bank of templates (with reduced density and sampling rate) in the first stage, which are then followed up in the second stage with the usual template bank (with optimal density and sampling rate) but only in the neighborhood of the parameter space of the candidate events identified in the first stage. Although the hierarchical search was demonstrated to speed up the standard PyCBC analysis by more than a factor of 20 in a previous work~\cite{kanchan_hierarchical}, assigning statistical significance to detected signals, especially in the presence of non-Gaussian noise, was done in a heuristic way. In this article, we present a robust approach for background estimation in a two-stage hierarchical search. Our method models the distribution of background triggers obtained from time-shifted triggers in a two-detector network. This modeling precisely aligns with the background distribution across critical signal-to-noise ratios (SNRs), where distinguishing between signal and noise is particularly challenging. It incorporates a fitting procedure to extrapolate to higher detection statistic values. Through an extensive injection campaign for a population of simulated signals on real data, we test the effectiveness of our background estimation approach. We find that our method achieves a sensitive volume-time product comparable to that of the standard two-detector PyCBC search. This equivalence holds for an inverse false alarm rate of 10 years by a factor ranging from $0.96\pm0.144$ to $1.02\pm0.145$ within the chirp mass range of approximately $1.4$ to $10~\text{M}_\odot$. Our methodology accomplishes this while substantially reducing the overall computational cost of the analysis. Specifically, our pipeline exhibits a remarkable speed-up, nearly 13 times faster than PyCBC analysis, including background estimation.
\end{abstract}

\maketitle

\section{\label{sec:level1}Introduction\protect}
Gravitational Wave (GW) astronomy has experienced impressive development since the landmark discovery of the first GW signal from the merger of the binary black hole (BBH), GW150914~\cite{gw150914}. Since then, the LIGO Scientific, Virgo, and KAGRA (LVK) collaboration has compiled an impressive catalog of nearly 90 GW sources~\cite{gwtc3_paper} comprising numerous binary black holes, two binary neutron stars (BNS)~\cite{gw170817_event,gw190425}, and two neutron star-black holes (NSBH)~\cite{nsbh_2021}. A majority of these detections have been possible through offline search pipelines~\cite{usman_2016,gareth,mbta1,mbta2,gstlal1,gstlal2,gstlal3,spiir} which primarily use matched filtering~\cite{svd,svd-sathya,svd-schutz,sathya-owen}, a model-based search technique that cross-correlates data from Advanced LIGO~\cite{asi} and Advanced Virgo~\cite{advancedvirgo} with a bank of modeled signals called \textit{templates}.%

In the current state-of-the-art, the search is mostly conducted for quadrupolar GW signals originating from mergers of compact binaries with quasi-circular orbits. With recent advancements in ground-based GW detectors, the ongoing upgrades in KAGRA~\cite{kagra_nature}, and the establishment of proposed detectors such as LIGO-Aundha~\cite{Saleem_2022_ligoindia} and third-generation detectors~\cite{3d_detectors}, the future holds promise for many more discoveries. It may even become possible to detect GW signals from binary systems, where parameter spaces expand by several orders of magnitude. This includes scenarios such as binary systems with orbital precession~\cite{ianharry_precessing_search} and eccentricity, as well as cases where the density of templates for spin-aligned systems increases, such as in subsolar mass compact binaries.%

Currently, such searches are either deferred or conducted within highly restrictive parameter spaces owing to computational limitations. This limitation arises because of the computational cost of matched filtering that scales with the template length, number of templates used, and dimensionality of the parameter space searched in the year-observed data. Consequently, there is an urgent demand for faster search pipelines that efficiently identify GW signals using these highly sensitive detectors. In this context, the application of hierarchical search methodology~\cite{bhooshan_hierarchical, rahul_nitz_hierarchical} has emerged as a promising approach for detecting GW signals.%

The hierarchical search method uses multiple banks of varying densities to matched filter detector's data. Initially proposed by~\citet{mohantysvd96}, matched filtering was performed by hierarchically searching over the chirp mass of nonspinning binary systems using Newtonian waveforms in stationary Gaussian noise. This approach was extended in subsequent studies to incorporate hierarchy over component masses using post-Newtonian (1.5PN) waveforms in~\citet{mohantysvd98} and over three parameters, component masses, and time of coalescence ($t_c$), in~\citet{anadsensvd2002,anandsensvd2003} using second-order post-Newtonian waveforms. It was also discovered that reducing the data sampling rate during the initial stage of the hierarchy improves the performance. %

~\citet{bhooshan_hierarchical} achieved a significant improvement in the hierarchical search method. Their work expanded the methodology to include all intrinsic parameters, such as binary component masses and spins and introduced a two-detector coincidence analysis into the algorithm. This advancement resulted in a remarkable speed-up of more than one order of magnitude in the Gaussian noise compared to traditional analysis techniques. Further exploration and development of this approach were conducted in~\citet{kanchan_hierarchical}, where a search pipeline was established and applied to data from the first two observing runs (O1 and O2) of Advanced LIGO. The study demonstrated the effectiveness of the hierarchical search method in successfully detecting all GW events previously detected by PyCBC search and reported in~\citet{gwtc1}. Notably, the hierarchical search significantly reduced the matched filtering computation by approximately 20-fold compared to the standard PyCBC (\textit{flat}) search method.%

The hierarchical search methodology described in \citet{kanchan_hierarchical} involved a two-stage matched filtering search within the PyCBC framework. In the first stage, a coarse search is performed on the data sampled at a low frequency using a sparsely sampled template bank called \textit{coarse bank}, generated using the hybrid geometric-random method~\cite{soumen,soumen2}. This enables faster matched-filtering operations, thereby facilitating a rapid parameter space scan. The objective of the coarse search is to identify potential coincident triggers that may be GW signals for the detector network. In the second stage, a more refined search is performed in the vicinity or neighborhood (\textit{nbhd}) of the coincident triggers identified during the coarse search. At this stage, matched filtering is executed at a higher sampling rate, enabling a meticulous investigation within a limited region of the parameter space. This approach maximizes the likelihood of detecting true GW signals, as the finer search focuses on regions of the parameter space where these signals are more likely to manifest. %

The two-stage hierarchical search strategy significantly reduces the computational cost of matched filtering. However, accurately assigning statistical significance to detected signals, particularly in real data containing non-Gaussian noise transients, remains challenging for this method. Specifically, estimating a GW event's false alarm rate (FAR) is not straightforward and requires careful consideration.%

The statistical significance of a GW event detected in one or more detectors is described by its FAR, which measures the likelihood of the event being a noise event rather than a true GW signal. There are various methods to estimate the FAR of a detected candidate. One commonly used approach involves artificially introducing time shifts to the data from one detector and searching for triggers coincident in time and the template parameters with the data from another detector~\cite{usman_2016,pycbclive_nitz}. Coincident triggers between detectors are recorded by repeating this process with many time shifts. Subsequently, FAR is determined for detected signals based on the number of noise coincidences that exceed a certain statistical threshold within a given background time.

Ideally, one can utilize the time-shift technique to estimate the background and determine the statistical significance of GW signals in the hierarchical search method. However, this technique may not achieve optimal effectiveness owing to two main factors. First, there is the potential issue of a biased background when the second search is conducted solely on zero-lag (\textit{foreground}) coincident triggers obtained from the first-stage search. This problem arises from insufficient background triggers, leading to a biased estimation of the GW event's FAR values detected in the second stage. Second, there is a tradeoff in computational advantages if all coincident triggers from the time-shift analysis of the first-stage search are followed for the second-stage search. While constructing a background by the time-shifting method, a larger number of triggers in the second stage may resolve the bias in FARs; it also increases the number of matched filtering operations, subsequently reducing the computational efficiency of the search.%

~\citet{bhooshan_hierarchical} addressed the issue of biased background estimation by proposing a heuristic approach that involves assigning significance using a scaled coarse background. This method was later demonstrated in~\citet{kanchan_hierarchical}, showing that a scaled background could reasonably approximate the significance of GW events. This approach lacks a concrete foundational basis for background estimation.%

In this work, we introduce a novel and robust approach to estimating the background in a two-stage hierarchical search. Our methodology empirically models the noise distribution obtained in the second stage of the search, particularly after implementing the time-shifting method. The procedure goes as follows:%

First, we conduct a coarse search to identify coincident triggers, encompassing foreground and background events as potential candidates for GW events. To mitigate excessive background triggers resulting from lower thresholds on single-detector statistics, we strategically reduce the number of time shifts in the first stage while maintaining a manageable increase in computational cost for matched filtering. The choice of time shift number ($\Nsh$) is determined proportionally to the computational cost ratio between the two stages, resulting in a practical reduction without compromising the overall computational gain. For our analysis, we select $\Nsh \gtrsim 148$ with a corresponding time shift of $\Tsh \sim 5,000$ s, applied to approximately 8.8 days of coincident data from LIGO's third observing run (O3). However, these values may be adaptable based on specific datasets. Our primary emphasis lies in identifying an optimal time shift value that yields a smooth background curve.%

In the second stage, we delve into the matched filtering process within the nbhds of the previously identified candidates from the coarse search. Employing matched filtering with templates in these nbhds, we collect resultant triggers for further analysis, focusing on evaluating GW signal significance. This assessment involves constructing a background distribution through time-shifting the data at intervals of 5,000 s. To enhance the accuracy of the noise distribution tail, we employ an empirical modeling technique. Specifically, we model the tail as a logarithmically decreasing curve with respect to ranking statistics. 

We test the effectiveness of our background estimation approach with an extensive injection campaign. Our proposed background estimation method gives a comparable sensitivity with the two-detector PyCBC search for binaries in the low-chirp mass region while significantly reducing the computational costs associated with matched filtering operations. Notably, our pipeline demonstrates an impressive speed-up, performing nearly 13 times faster than the PyCBC search.%

The paper is organized as follows: In Sec.~\ref{sec:review_flat}, we provide a comprehensive review of the search methodology employed by the flat search. We revisit the definitions of \textit{one-step}\footnote{matched-filtering performed over data sampled at a unique frequency and a template bank.} matched filtering technique, delve into the data acquisition process and data-quality checks, explain the derivation of single-detector and ranking statistics pertinent to the two-detector search, and briefly describe their approach to estimating the FAR. These definitions serve as the foundation for our subsequent discussions in Sec.~\ref{sec:hierarchicalsearch}, where we review the hierarchical search method for a two-detector network. In Sec.~\ref{sec:far_hierarchical}, we present our new method for background estimation designed explicitly for the hierarchical search. In Sec.~\ref{sec:application}, we examine the robustness of our background estimation approach by applying it to real data and comparing its consistency with the flat search. Furthermore, in Sec.~\ref{sec:sensitivity_comparison}, we compare the sensitivity of our search pipeline with that of the flat search. In Sec.~\ref{sec:computationalcost}, we evaluate the computational efficiency of our pipeline. Finally, we conclude our findings and draw inferences in Sec.~\ref{sec:conclusion}.

\section{\label{sec:review_flat}Review of existing two-detector Flat search}
The search for GW signals embedded in the noise $n(t)$ in a two-detector configuration is performed via a matched filtering technique using the PyCBC-toolkit~\cite{usman_2016}. This technique involves cross-correlating the output $s(t)$ of an interferometer with a modeled waveform known as template $h(t_c,\phi_c,\vec{\theta})$ in the frequency domain. The correlation is performed for different values of the coalescence time ($t_c$) and phase ($\phi_c$), maximizing the matched-filter SNR $\rho(t)$ given the source parameters\footnote{$\theta$ includes the component masses ($m_1,m_2$) and dimensionless spin vectors ($s_{1z}, s_{2z}$) of a spin-aligned binary system.} ($\vec{\theta}$). Mathematically, this can be expressed as  
\begin{equation}\label{subeq:snr}
\rho \left( t_c; {\vec{\theta}} \right) \equiv 
    \left| \left( s, (1+i)h({ t_c,\phi_c = 0,\vec{\theta}})\right) \right| \,.
\end{equation}
In Eq.~\ref{subeq:snr}, $(.,.)$ represents the scalar product between any two data time series $x(t)$ and $y(t)$ weighted by the one-sided power spectral density (PSD) $S_n(f)$ of the interferometer. The scalar product is defined as 
\begin{equation}\label{subeq:scalarprod}
(x, y) := 4 ~\mathcal{R}\bigg\{
    \int^{f_{high}}_{f_{low}} \frac{\tilde{x}(f) \tilde{y}^{*}(f)}{S_n{(f)}} \,df \bigg\}   \,.
\end{equation} 
Here, $\tilde{x}(f)$ and $\tilde{y}(f)$ denote the Fourier transforms of $x(t)$ and $y(t)$ respectively, and $\mathcal{R}$ represents the real part of the complex number. The integration is performed over a frequency range of $f_{low}$ to $f_{high}$.%

As the incoming signal parameters are not known in advance, matched-filter statistics are computed on data sampled at 2048 Hz using a flat template bank (see Table~\ref{table:table1}). If the peak of the matched-filter SNR $\rho$ is greater than or equal to a predefined threshold $\rho_{thr}$ ($\rho_{thr}=4$), a \textit{trigger} is stored for further analysis. However, these triggers can be generated due to non-Gaussian transients in the data, resulting in many false alarms. Therefore, the triggers are down-weighted using their chi-square ($\chi^2_{r}$) ~\cite{bruceallen} and sine-Gaussian chi-square ($\chi_{sg}^2$) values~\cite{nitzsgveto}. Using these chi-squares, single-detector statistics are computed, which are, 
\begin{align}
\label{subeq:newsnrsgveto}
\hat{\rho} = \begin{cases}
\tilde{\rho}~{(\chi^2_{sg})}^{-1/2} & \text{if $\chi^2_{sg} \ge 6$}, \\
\tilde{\rho} & \text{otherwise},
\end{cases}
\end{align}
where $\tilde{\rho}$, 
\begin{align}
\label{subeq:newsnr}
\tilde{\rho} = \begin{cases}
\frac{\rho}{[(1+(\chi_r^2)
^{3})/2]^{1/6}} & \text{if $\chi_r^2 \ge 1$}, \\
\rho & \text{otherwise}.
\end{cases}
\end{align}

After re-weighting, the surviving triggers are subjected to a coincidence test to identify instances of time coincidence with a high value of ranking statistics ($\Lambda$)~\cite{nitzphasetd} within the two-detector network. This statistic is given by 
\begin{equation}
    \Lambda = \frac{p(\vec{\kappa} | S)}{p(\vec{\kappa} | N)} = \frac{p(\vec{\kappa} | S)}{r^{HL}_{\theta}~p(\theta,\delta t_{c,HL} , \delta \phi_{c,HL} | N)}, 
\end{equation}\label{eq:nitzphasetd}
is the ratio of coincident event rate densities due to signal $p(\vec{\kappa} | S)$ and noise $p(\vec{\kappa} | N)$ for a coincident trigger's template parameters $\vec{\kappa} = \{\hat{\rho}_{H},~\hat{\rho}_{L},~\chi_{sg,H}^{2},~\chi_{sg,L}^{2},~\delta t_{c,HL}, \delta \phi_{c,HL},~\theta\}$. %

Due to non-Gaussian features like glitches in the data, many loud coincidences may occur, posing a challenge in distinguishing true GW signals from noise and assessing their significance. Therefore, to measure the significance, FAR is computed by performing time shifts to the triggers by 0.1 s and recalculating the ranking statistics between the detectors. By generating a background of false coincidences ($n_b$) through multiple time shifts, the generated distribution is used to compute the FAR~\cite{usman_2016} as follows:

\begin{equation}\label{subeq:far}
\text{FAR} = \frac{\left(1 + n_b(\Lambda_{b} \ge \Lambda^*_{f})\right)}{T_b},
\end{equation}\label{far_usman}
where $T_b$ represents the observation time for the background estimation, $\Lambda_b$ denotes the coincident statistic values of background triggers, and $\Lambda^*_f$ for the foreground trigger that may or may not define the real GW event.

\section{\label{sec:hierarchicalsearch}Review of Two-stage Hierarchical search}
The hierarchical search involved a matched filtering search in two stages, as outlined in~\cite{kanchan_hierarchical}. In the first stage, data segments sampled at 512 Hz are filtered using the templates from a coarse bank specified in Table~\ref{table:table1}. Triggers with $\rho$ and $\tilde{\rho}$ values exceeding 3.5 are collected for each detector in the network. For templates with a total mass greater than $30~\text{M}_\odot$, the triggers are further re-weighted using $\chi^2_{sg}$ to mitigate the impact of short-duration glitches on the noise. The selected triggers undergo coincidence testing where ranking statistics, denoted by $\Lambda_1$, are computed for the foreground and background triggers.%

\begin{table}[ht]
\centering
\caption{Summary of minimal match values and parameter ranges for coarse and flat banks. The banks are designed to search for redshifted total mass in the range [2, 500] M$_{\odot}$, dimensionless spin parameters for black holes in [-0.99, 0.99], and neutron stars in [-0.05, 0.05].}
\vspace{0.2cm}
\begin{tabular}{
l@{\hspace{40pt}}
c@{\hspace{40pt}}
c@{\hspace{5pt}}}
\hline
\hline
\addlinespace
Bank & Templates & Minimal match (MM)  \\
\addlinespace
\hline
\addlinespace
Coarse & 85,080 & 0.90 \\
Flat & 428,725 & 0.97  \\
\addlinespace
\hline
\hline
\end{tabular}
\label{table:table1}
\end{table}

Unlike in the flat search, a background is constructed in this stage by performing multiple time shifts with an interval of $\Tsh = 5,000$ s. This approach serves two primary purposes. Firstly, it enables the examination of all coincident triggers, encompassing both foreground and background triggers, during the second stage search, where a finer search is conducted in the vicinity of the followed-up trigger's template parameter space. This comprehensive analysis predominantly accounts for noise coincidences in the second-stage search, which were not previously accounted for in~\citet{kanchan_hierarchical}. Secondly, conducting time shifts of $\Tsh = 5,000$ s is more computationally efficient. This approach entails performing a finer search on fewer coincident triggers identified in the first stage, thereby avoiding a significant increase in the computational cost of matched filtering in the second stage.%


\begin{figure}[ht!]
    \centering
    \includegraphics[scale=.3]{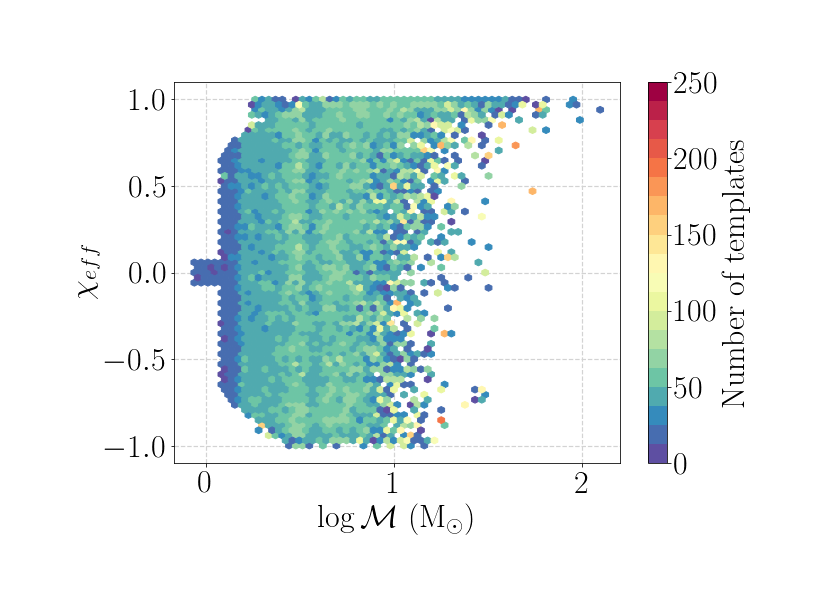}
    \caption{Plot depicting the distribution of coarse templates in the chirp mass ($\mathcal{M}$) - effective spin ($\chi_{\text{eff}}$) plane. The color bar represents the expected number of templates in the nbhd of each coarse template parameter.}
    \label{fig:nbhd_bank}
\end{figure}

The second stage search is conducted for data segments containing triggers with $\Lambda_1 \ge 7$. In this stage, these data segments are re-sampled at 2048 Hz and undergo matched filtering with templates in the nbhd of followed-up coarse template parameters. Each coarse template consists of nearly tens or hundreds of templates in its respective nbhds as depicted in Fig.~\ref{fig:nbhd_bank}. When the second stage search is conducted, a union of these nbhds is constructed for matched filtering of the data segments. At this step, triggers with $\rho$ and $\tilde{\rho}$ above 4 are collected for further analysis. These triggers are again down-weighted with $\chi^2$ and $\chi^2_{sg}$ like in the first stage search. The surviving triggers then undergo a coincidence test, where coincidence statistics $\Lambda_2$ is calculated. This statistic explicitly represents $\Lambda$ associated with the second stage search.%

\section{\label{sec:far_hierarchical} A direct method for estimating the FAR in the hierarchical search} 
The distribution of coincident triggers obtained in the second stage from time-shifted analysis is used for estimating FAR for GW events. To calculate FAR, we employ a hybrid approach that combines the time-shifted background with a fitting process applied to the tail of the distribution. This fitting process involves using a falling exponential model to account for true noise triggers. The application of the exponential-fitting procedure has previously been explored in estimating significance for single-detector events, as demonstrated in~\citet{gareth_singledet_2023}. In the present study, we leverage this method to address the distribution of time-shifted coincidence triggers. The procedural approach is as follows:

First, we apply multiple time shifts with an interval of 5,000 s to the triggers. This leads recalculation the quantity $\Lambda_2$. By segregating the foreground coincidences from the time-shifted ones, we obtain a distribution consisting solely of the true noise triggers.%

Notably, a time-shift interval of 5,000 s corresponds roughly to the generation of one year's worth of background data. In contrast, the background generated by the flat search, using a time-shift interval of 0.1 s, spans a considerably longer duration. To ensure a fair comparison, we generate a comparable amount of background by extrapolating the tail of the cumulative distribution of $\Lambda_2$. This extrapolation allows us to estimate the background level for a time duration equivalent to the 0.1 s time-shift interval, enabling a meaningful assessment of the FAR estimation.%

\begin{figure}[ht!]
    \centering
    \includegraphics[scale=.21]{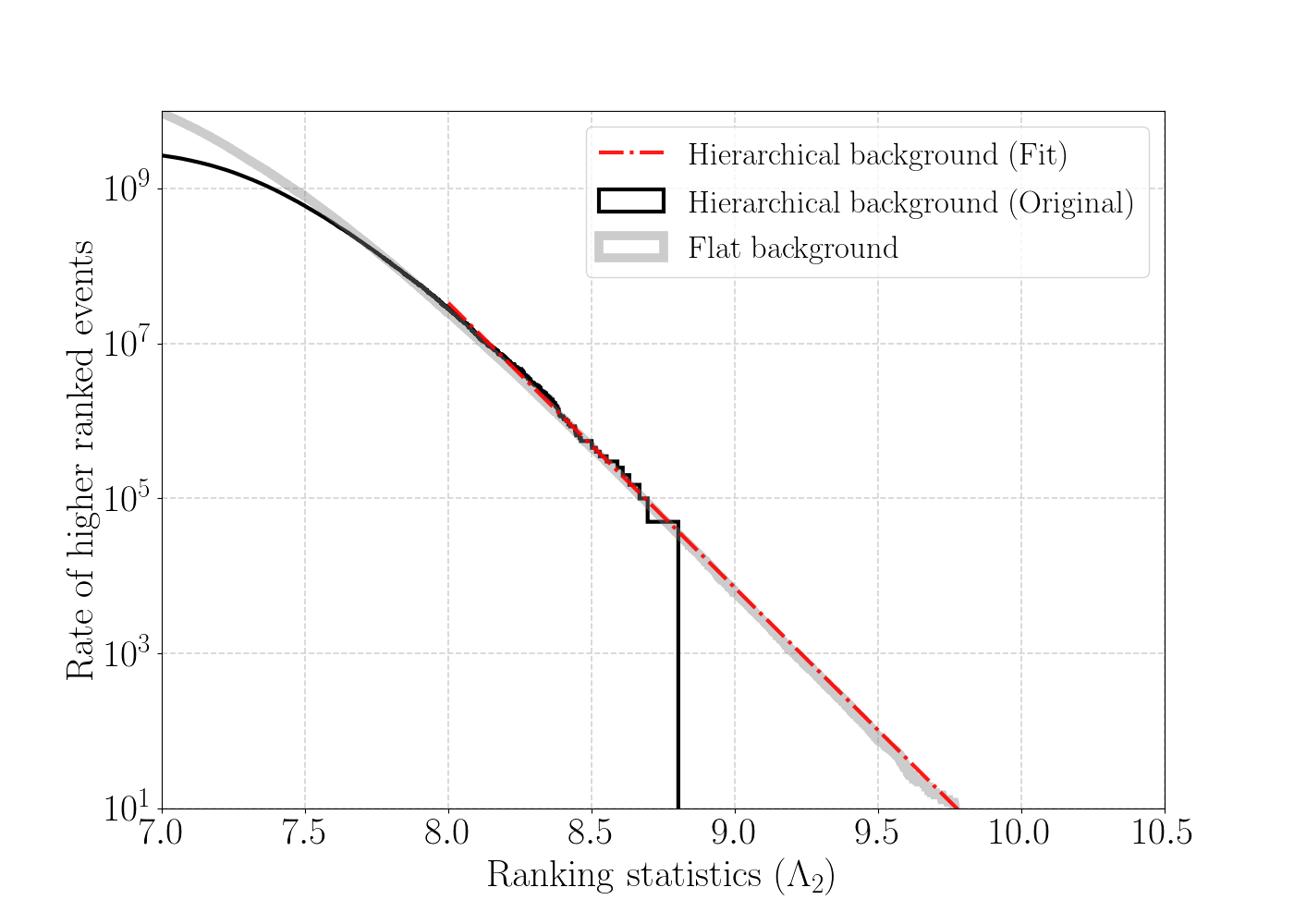}
    \caption{Plot depicting the rate of higher-ranked background events (excluding the foreground) versus ranking statistics for both the flat and hierarchical searches. The background events derived from a time-shift interval of 0.1 in the flat search are depicted in grey, while those from a 5,000 s interval in the hierarchical search are shown in black. A linear fit to the hierarchical background (red) is superimposed to demonstrate that the cumulative count of events decreases linearly on a logarithmic scale with the ranking statistics.}
    \label{fig:fit_o3_cumulative_counts}
\end{figure} 

Before proceeding with the extrapolation, we appropriately scale the background distribution by assuming that the number of noise coincidences in the second stage increases by a factor determined by the ratio of the time-shift intervals used in the hierarchical search (5,000 s) and the flat search (0.1 s). This scaling ensures that the cumulative distribution of the hierarchical background matches that of the flat background, as shown in Fig.~\ref{fig:fit_o3_cumulative_counts}. Note that this scaling differs greatly from what was done in \citet{kanchan_hierarchical}. In the present work, the scaling is in total data volume, which is justifiable, as false alarm distribution is broadly independent of the length of the data. This is quite different from scaling the coarse-bank background to match that for the flat-bank using sampling rate and template number argument, as there is no clear proof that the scaling should be linear at all values of the detection statistic for these parameters.%

Figure~\ref{fig:fit_o3_cumulative_counts} also shows that due to the limited number of time shifts employed in the second stage search, there is a scarcity of coincident triggers. Consequently, the maximum value attained by $\Lambda_2$ is approximately $8.8$, which may not be sufficient for estimating the FAR. However, the background distribution obtained the time-shifting method, exhibits an approximately linear trend on a logarithmic scale as $\Lambda_2$ increases. This observation suggests using linear regression methods to extrapolate the curve and estimate trigger values at higher values of $\Lambda_2$. %

Accordingly, we perform a linear regression on the cumulative distribution of the number of background coincident triggers represented on a logarithmic scale. Mathematically, for high $\Lambda_2 \gtrsim 8$, the linear regression equation can be written as,
\begin{equation} 
   \log(n_{b,2}) = \alpha + \beta \Lambda_2, 
\end{equation}
where $\alpha$ and $\beta$ are the fit parameters and $n_{b,2}$ is the number of background triggers in the second stage exceeding $\Lambda_2$. The linear regression assumes that the tail of the true noise coincidences follows a power law that linearly decreases as the value of the statistics increases. This argument holds even when more time slides are employed when following up coincident triggers from the first stage search. However, this approach may increase the number of matched filtering operations and reduce the computational advantages of hierarchical search over the flat search. 

Once we have obtained the fitted noise distribution, we compute FAR for the foreground events detected in the second stage of the search. This involves utilizing Eq.~(\ref{subeq:far}) and the fitted distribution to determine the number of background coincidences that have a $\Lambda_{2}$ value greater than or equal to that of the foreground triggers' statistics.%

\begin{table*}[ht!]
\centering
\caption{Results from a two-detector analysis over data duration from May 12 to May 21, 2019, using flat and hierarchical searches. The events detected from both searches are arranged in descending order of network SNR values up to a FAR value of 1 per year. The FAR estimates in the case of the flat search are obtained using the time-shift method, while the FARs in the hierarchical search are obtained using our proposed method described in Sec.~\ref{sec:far_hierarchical}. The re-detection of these events is consistent with the third gravitational wave transient catalog (GWTC-3)~\cite{gwtc3_paper}.}
\vspace{0.2cm}
\begin{tabular}{
l@{\hspace{20pt}}
c@{\hspace{10pt}}c@{\hspace{10pt}}c@{\hspace{20pt}}c@{\hspace{10pt}}
c@{\hspace{10pt}}c@{\hspace{3pt}}}
\hline
\hline
\addlinespace
 & \multicolumn{2}{c}{Flat} & \multicolumn{3}{c}{Hierarchical} \\
 \addlinespace
\cline{2-3} \cline{5-6}
\addlinespace
Event & FAR (yr$^{-1}$) & Network SNR & & FAR (yr$^{-1}$) & Network SNR \\
\addlinespace
\hline
\addlinespace
GW$150921\_074359$ & $<1.76\times 10^{-5}$  & 24.03 & & $<1.76\times 10^{-5}$ & 23.31 \\
GW$190519\_153544$ & $<1.76\times 10^{-5}$ & 13.19 & &  $<3.53\times 10^{-5}$ & 12.55 \\
GW$190513\_205428$ & $0.37\times10^{-3}$ & 11.61 & & $0.19\times10^{-3}$ & 11.79 \\
GW$190517\_055101$ & $0.09\times10^{-1}$ & 10.13 & & $0.34\times10^{-1}$ & 10.45 \\
\addlinespace
\hline
\hline
\end{tabular}
\label{table:events}
\end{table*}

\section{\label{sec:application}Application to Real Data}
We now employ a hierarchical search pipeline on real data, as elaborated in Sec.~\ref{sec:hierarchicalsearch}. We apply the methodology outlined in Sec.~\ref{sec:far_hierarchical} to calculate FARs for the detected events in our analysis. For comparison, we will utilize a two-detector PyCBC-broad search pipeline.%

To facilitate this investigation, we analyzed approximately 8.8 days of data extracted from May 12 to May 21, 2019, during the third observing run of the twin LIGO detectors. We used the strain data obtained from the Data Quality Segment Database (DQSEGDB)~\cite{dqsegdb} for the detectors at Hanford and Livingston. This data encompasses a mixture of Gaussian and non-Gaussian noise features. The presence of the latter, often attributed to diverse instrumental artifacts, requires their removal before the matched-filtering step. Therefore, the timestamps displaying suboptimal quality or instances of data unavailability were removed using Category 1 and Category 2 vetoes. Moreover, transient short-duration artifacts in the data were eliminated using a gating method~\cite{gating_Davis_2021}. Once the timestamps unsuitable for astrophysical searches were removed, we ran both search pipelines to initiate the analysis.%

In the analysis, we used the pre-computed flat bank for flat search and coarse and nbhd banks for hierarchical search, as elaborated upon in~\citet{kanchan_hierarchical}. These sets of banks were specifically designed to explore the parameter spaces of the BBH, BNS, and NSBH systems. Further, the banks were tailored to cover a wide range of total masses, from 2 to 500 M$_\odot$, as well as component spins within the respective ranges of 0 to 0.9 for black holes and 0 to 0.05 for neutron stars. Spin precession was not considered in this study.%

\subsection{\label{sec:results}Results}
\begin{figure}[ht!]
    \centering
    \includegraphics[scale=.25]{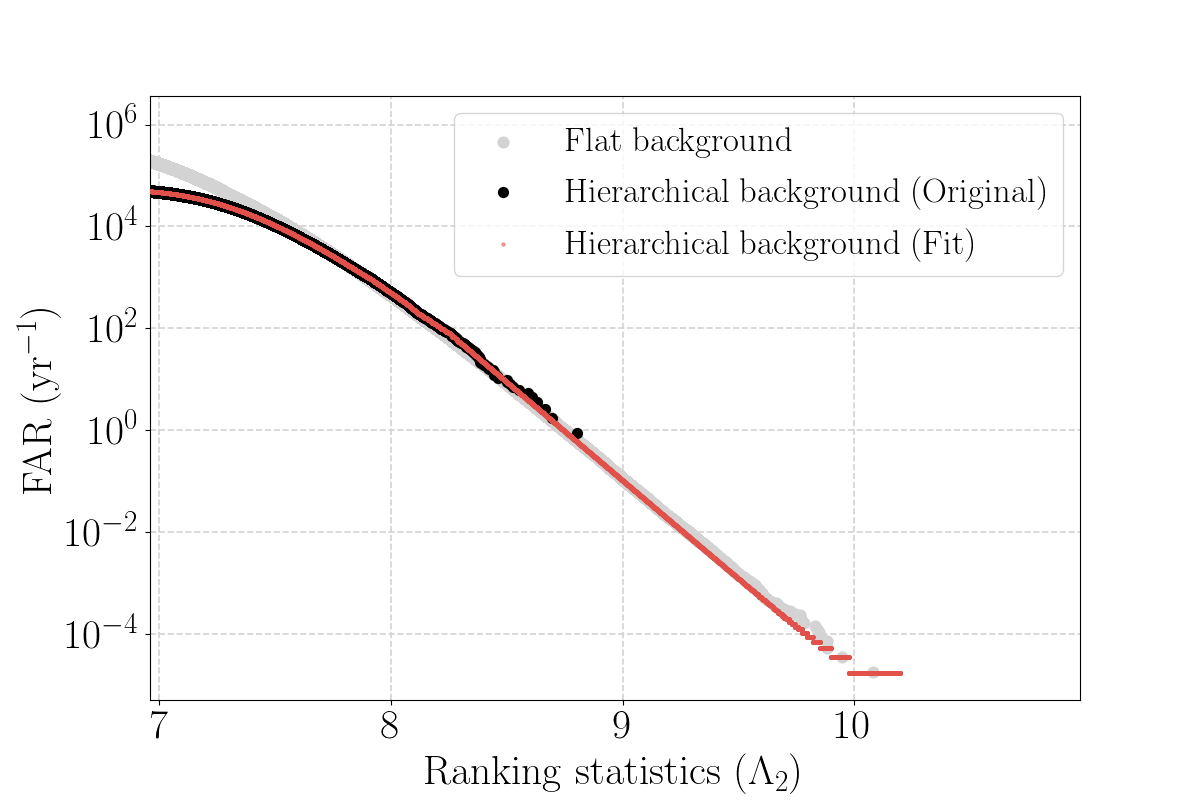}
    \caption{A Comparison plot of the FAR as a function of ranking statistics of the background triggers obtained in flat and hierarchical search methods. The FAR distribution with and without exponential fit to the hierarchical background distribution is shown in black and red, respectively. The FAR distribution for background from the flat search is overlaid (gray) to demonstrate that the linear-fit approximation adequately matches the FAR values between the two search methods.}
    \label{fig:plot_far_o3}
\end{figure}

The analysis resulted in the identification of numerous foreground and background coincidences. We assessed the significance of foreground triggers using the method outlined in Sec.~\ref{sec:far_hierarchical}. As depicted in Fig.~\ref{fig:plot_far_o3}, the FARs of the background triggers obtained from the background fit closely match those derived from the traditional time-shifting method in the hierarchical search up to a threshold of approximately $\Lambda_{2}\sim8.8$. Furthermore, the fit aligns well with the curve obtained from the flat search, wherein a larger number of time shifts, equivalent to an interval of 0.1 s, are employed. This outcome demonstrates the effectiveness of our proposed approach for background estimation, as elaborated in Sec.~\ref{sec:far_hierarchical}.%

\begin{figure}[ht!]
    \centering
    \includegraphics[scale=0.23]{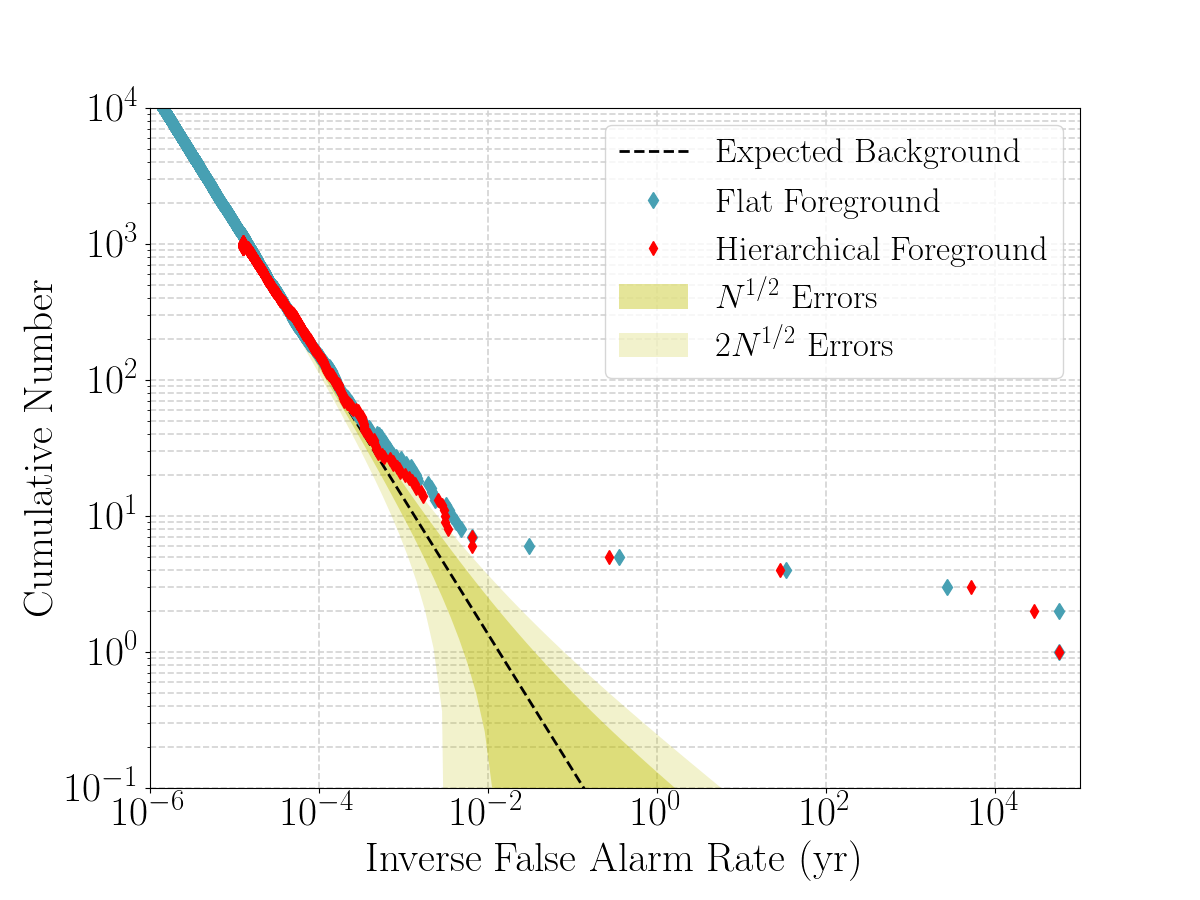}
    \caption{Comparison of the cumulative histogram of the inverse false alarm rate ($1/\text{FAR}$) for the foreground triggers obtained in the flat (blue) and hierarchical (red) searches. The dashed line represents the expected background for a given observation time $T$, while the yellow-shaded regions represent Poisson errors estimated for the flat search.}
    \label{fig:comparison_inverse_far}
\end{figure}

We also examined the significance of foreground events identified in our analysis. Table~\ref{table:events} summarises our findings from the two searches. Remarkably, our method identified all events characterized by high network SNRs with consistent FAR values within a factor of a few compared to those obtained from the flat search. We were able to recover two previously detected GW events-- GW$150921\_074359$ and GW$190519\_153544$, as cataloged in~\citet{gwtc3_paper}. Additionally, we identified two other low-significance events, GW$190513\_205428$ and GW$190517\_055101$, also reported in the GWTC-3. These findings are also illustrated in Fig.~\ref{fig:comparison_inverse_far}, where we present a comparison of the Inverse False Alarm Rates (IFARs) for the recovered events in both hierarchical and flat search methodologies. The results indicate that our approach effectively estimates the IFARs  for significant events detected in the O3 data. Consequently, this enhances the reliability and validity of our background estimation methodology.%

\begin{figure*} 
    \centering
    \includegraphics[scale=0.17
]{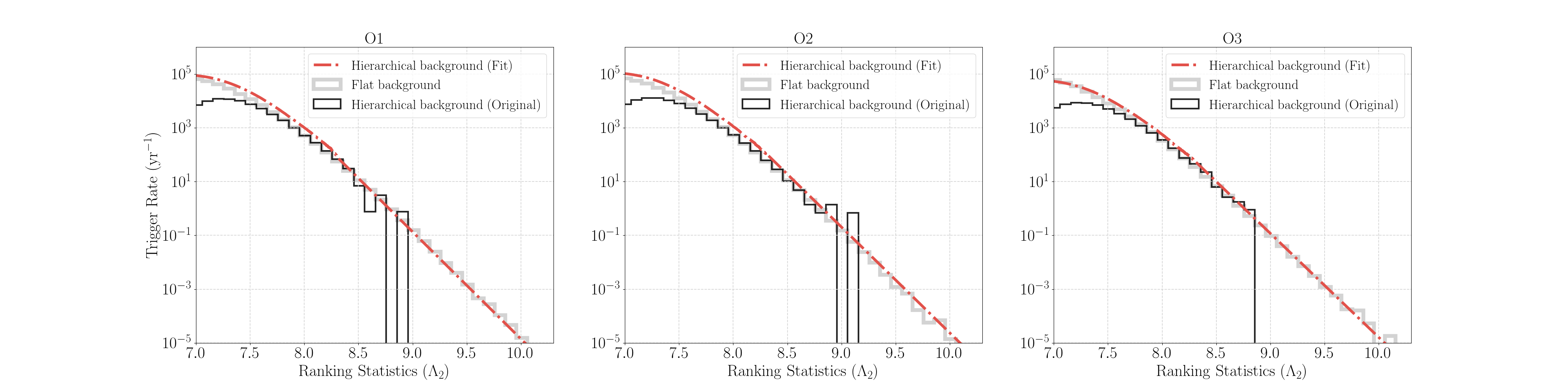}
    \caption{Comparison of the histograms of background trigger rates (excluding the foreground) versus ranking statistics for O1, O2, and O3 data. The black curve represents the true background obtained from time-shifted triggers across the detectors. The red curve represents the background obtained after fitting the true background trigger distribution. The background trigger rate from a flat search is overlaid in gray for comparison.}
    \label{fig:fit_o1_o2_o3}
\end{figure*}

The data quality employed in the analysis could affect the fitting procedure for acquiring the background. To assess the efficacy of our methodology across various instances of noise, we applied it to multiple datasets collected during the first two observing runs of the twin LIGO. Specifically, our analysis concentrated on approximately 5.3 days of concurrent data from O1 and 5.5 days from O2. Encouragingly, our approach consistently demonstrated a significant concurrence between the background trigger rates obtained through our proposed methodology and those derived from the flat search, as depicted in Fig. ~\ref{fig:fit_o1_o2_o3}. This further demonstrates the robustness of the background estimation approach.%

\subsection{\label{sec:sensitivity_comparison}Detection Efficiency: Validation of background estimation approach}
In the previous section, we identified similarities in the FARs computed for the events detected from the two search pipelines. This observation motivated us to assess the robustness of our background estimation approach using an extensive injection campaign. In this campaign, we initially inject simulated signals into the data and recover them using search pipelines at a specified FAR. We then compare the sensitivity of the search pipelines by computing the sensitive volume-time (VT) product.%

The VT product for a search pipeline measures the projected number of detectable signals originating from a population of binary systems that exceeds a predefined level of statistical significance. Assuming the binary merger rate remains constant, the ratio of the VT product between the two search pipelines provides insight into their relative sensitivities. Typically, this evaluation is carried out using Monte Carlo simulations~\cite{tiwari_vt}.%

To facilitate this study, we systematically injected various signals corresponding to the BBH, BNS, and NSBH mergers. We then observed their recovery using both flat and hierarchical searches, as described in Sec. ~\ref{sec:review_flat} and Sec. ~\ref{sec:hierarchicalsearch} respectively. Specifically, we targeted sources with intrinsic parameters as listed in Table~\ref{table:table_injection}. To simulate these sources, we generated non-precessing quasi-circular and quadrupolar GW signals using the \verb+SpinTaylorT4+~\cite{spintaylor} waveform model~\cite{spintaylor} for BNS sources and the \verb+SEOBNRv4_opt+~\cite{seobnr} waveform model for BBH and NSBH sources. These signals were injected uniformly across time intervals of 100 s within the data. For simplicity, we distributed these signals uniformly across inclination and chirp distances ranging from 5,000 to 400,000 Mpc.%

\begin{table}[ht!]
\caption{The table presents the parameter distributions and corresponding ranges for $m_{1}$, $m_{2}$ (redshifted component masses), and $\chi_{1}$ and $\chi_{2}$ (dimensionless effective spins) for each source category.}
\vspace{0.2cm}
\centering
\begin{tabular}{ l@{\hspace{20pt}}
c@{\hspace{15pt}} c@{\hspace{15pt}} r@{\hspace{5pt}}}
\hline\hline
\addlinespace
Source & Parameter & Distribution & Range \\
\addlinespace
\hline
\addlinespace
BBH & $m_{1}, m_{2}$ & Log-uniform & 2.5--150 M$_\odot$ \\
& $\chi_{1}$, $\chi_{2}$ & Uniform & 0--0.9 \\
\addlinespace
\hline
\addlinespace
BNS & $m_{1}, m_{2}$ & Uniform & 1--2.5 M$_\odot$ \\
& $\chi_{1}$, $\chi_{2}$ & Uniform & 0--0.4 \\
\addlinespace
\hline
\addlinespace
NSBH & $m_{1}$ & Log-uniform & 2.5--97.5 M$_\odot$ \\
& $m_{2}$ & Log-uniform & 1--2.5 M$_\odot$ \\
& $\chi_{1}$ & Uniform & 0--0.9 \\
& $\chi_{2}$ & Uniform & 0--0.4 \\
\addlinespace
\hline\hline
\end{tabular}
\label{table:table_injection}
\end{table}

\begin{figure*}
    \centering
    \includegraphics[scale=0.5]{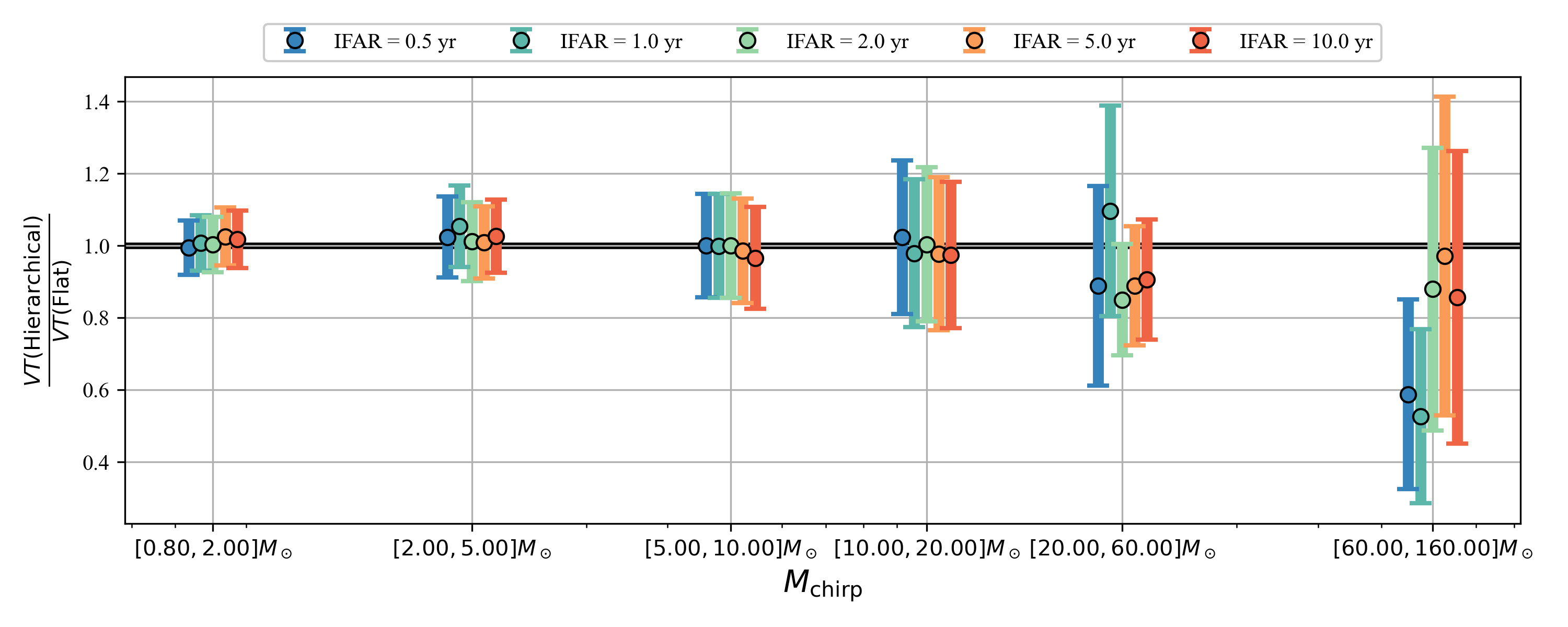}
    \caption{Plot depicting Volume-time (VT) ratio between hierarchical and flat search.}
    \label{fig:vtplot}
\end{figure*}

The recovered injections were observed and subjected to VT product computations to compare the sensitivities of the pipelines. Our investigation yielded compelling results, as shown in Fig. ~\ref{fig:vtplot}. As can be seen, the VT ratios between the two pipelines for low chirp masses and across all the IFAR bins are comparable. Specifically, the values fluctuate between $0.96\pm0.144$ and $1.02\pm0.145$ for chirp masses $\sim 1.4-10~\text{M}_\odot$ at an IFAR of 10 years. This result indicates that the hierarchical search's sensitivity is comparable to that of the flat search, particularly for binary systems characterized by low-chirp masses. However, the sensitivity of hierarchical search decreases slightly for chirp masses exceeding $60~\text{M}{\odot}$. This reduction is primarily attributed to the lower density of coarse templates within the higher chirp mass range and the domain of low effective spins, as shown in Fig. ~\ref{fig:nbhd_bank}. As a result, the effective number of nbhd templates in this specific parameter region was notably decreased, leading to a reduced count in recovering the injected signals.%

\subsection{\label{sec:computationalcost}Computation Efficiency} 
We demonstrated that our background computation exhibits remarkable signal detection capabilities while maintaining a search sensitivity comparable to the flat search in low-chirp mass regions. To further evaluate the performance of our search pipeline, we now turn our attention to its computational efficiency.%

The computational efficiency of a search pipeline relies heavily on the speed at which matched filtering operations, which inherently involve performing Fast Fourier Transforms (FFTs),  are executed. This aspect can be evaluated by directly utilizing the Cooley-Tukey algorithm~\cite{cooley-tukey} to compute the number of floating-point operations involved.%

Let us assume that the data segment is sampled at a frequency of $f$ Hz and has a $t_{\rm seg}$ s duration. In this case, the number of FFT operations can be expressed as $\kappa~\text{N}\log_{2}(\text{N})$, where $\text{N}=f~t_{\rm seg}$ and $\kappa$ is approximately a few. This operation is repeated for several templates denoted by $\text{N}_{\rm temp}$, resulting in an actual computational cost of approximately $\kappa \text{N}_{\rm temp}~\text{N}\log_{2}(\text{N})$. It means that the computational cost of matched filtering scales with the data sampling rate and number of templates. %

Ideally, the speed-up factor in the matched filtering computation between the flat and hierarchical search methods can be estimated by calculating the ratio of the total cost in the flat search to the combined total cost in both stages of the hierarchical search. This speed-up can be represented as
\begin{equation}
    \text{speed-up} \approx \frac{N_{\rm seg} N^{\rm flat}_{\rm temp}~O_{\rm flat}}{N_{\rm seg} N^{\rm stage1}_{\rm temp}~ O_{\rm coarse} +  N^{\rm stage2}_{\rm temp}~O_{\rm fine}}\,,
\end{equation}
where,
\begin{eqnarray*}
    O_{\rm flat} = \kappa f_{\rm flat}t_{\rm seg}~\log(f_{\rm flat}t_{\rm seg})\,,  \\
    O_{\rm coarse} = \kappa f_{\rm coarse}t_{\rm seg}~\log(f_{\rm coarse}t_{\rm seg}) \,, \\
    O_{\rm fine} = \gamma ~O_{\rm flat} \,.
\end{eqnarray*}
Here, $N_{\rm seg}$ represents the total number of overlapping segments analyzed. The data sampling rates are denoted by $f_{\rm flat}$ (2048 Hz) for the flat search and $f_{\rm coarse}$ (512 Hz) for the first stage of the hierarchical search. While the matched filtering cost dominates the analysis of a data segment with a large template bank, overhead costs are associated with each segment, for example, for loading and initial data processing, which are independent of the number of templates. The overhead cost per template can become comparable to or even greater than the matched filtering cost when the number of templates is small. In the above formula, factor $\gamma$ accounts for this.%

As mentioned earlier, we compared the matched filtering computational costs between the hierarchical and flat search pipelines using O3 data from the Hanford and Livingston detectors. For the analysis, the observed data was divided into 176 blocks from the Hanford detector and 197 blocks from the Livingston detector. Each of these blocks was further divided into overlapping segments of 512 s length.%

\begin{figure}[ht!]
    \centering
    \includegraphics[scale=0.25]{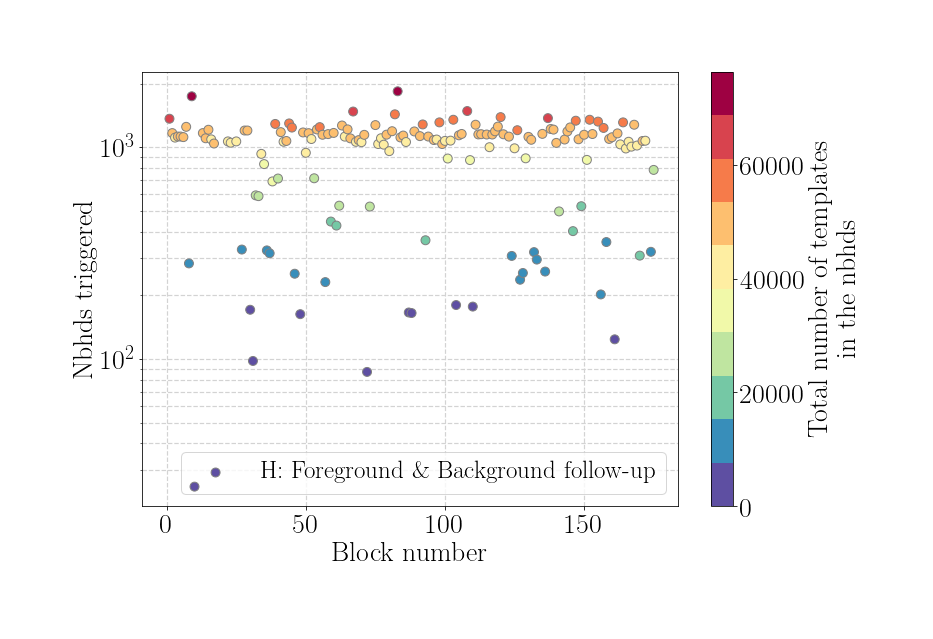}
    \includegraphics[scale=0.25]{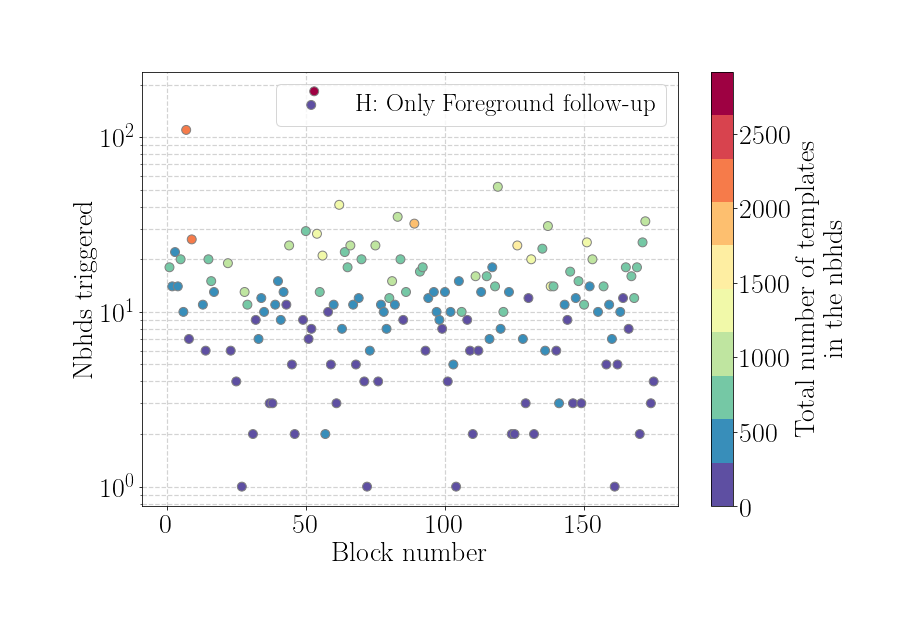}
    \caption{Plot depicts the number of nbhds triggered per block while performing the second-stage search for data from Hanford (H). The color bar represents the total number of templates in the nbhds triggered for a block. The number of nbhds triggered and the corresponding number of templates in the nbhds is larger when foreground and background triggers are followed up in the second stage (top panel). However, these numbers are lower when only the second search is performed on the foreground triggers alone (bottom panel).}
    \label{fig:segment_temp}
\end{figure}

In the flat search, each segment of the data block sampled at a rate of $f_{\rm flat}$ was filtered with $\text{N}^{\rm flat}_{\rm temp} = 428,725$ templates. However, the number of templates and data sampling rates varied between the two stages of the hierarchical search. In the first stage, the data segments were sampled at $f_{\rm coarse}$ and filtered with 85,080 templates denoted by $N^{\rm stage1}_{\rm temp}$. During the second stage of the search, despite the data segments being sampled at $f_{\rm flat}$ rates, the number of templates varied across segments, leading to fewer FFT operations than in the flat search. This variability in the number of templates per segment resulted from the number of templates in the union of the nbhds that were followed. Consequently, the number of templates fluctuates for each data block, as shown in Fig. ~\ref{fig:segment_temp}. %

If the number of templates used by each segment were large, $\gamma$ would be close to $1$. However, we find that for a typical data block, for the Hanford (Livingston) detector, the CPU time taken for flat search, with $428,725 \times 10$ (segments in the block) $ \sim 4$ million application of templates to data, was 66.52 hr (73.15 hr). In contrast, for the same block, in the second stage, with a total of 46,369 applications of templates to data, the CPU time taken was 1.88 hr (1.72 hr). If we compare the computation time per segment per template, one can see that the flat search ($\sim 0.05$ s per template per segment) was approximately three times less expensive than the nbhd search ($\sim 0.15$ s per template per segment), that is, $\gamma \sim 3$. We checked for different data blocks and $\gamma$ remained nearly the same for similar data block lengths.%

In our study, with $t_{\rm seg}$ set at 512 s and $N_{\rm seg}$ set at 1,493 for the Hanford detector (or 1,731 for Livingston), the total number of templates in the second stage of the hierarchical search was approximately 6,274,846 (or 6,235,197 for Livingston). The resulting speed-up in the matched-filtering computation was approximately 18.36 (or 18.38 for Livingston). Accounting for the overhead computation time per segment with $\gamma \sim 3$, the estimated speed-up, 13.58 for Hanford and 14.07 for Livingston, matches the actual CPU obtained by comparing the real CPU times.%

\begin{table}[ht!]
\centering
\caption{Table depicting the CPU core hour for the matched filtering jobs in flat and hierarchical searches. The numbers are estimated separately for the two detectors-- Hanford (H) and Livingston (L) in the analysis.}
\vspace{0.2cm}
\begin{tabular}{
l@{\hspace{20pt}} 
c@{\hspace{15pt}} c@{\hspace{2pt}}}
\hline
\hline
\addlinespace
Search pipeline & \multicolumn{2}{c}{CPU core hour} \\
\addlinespace
\cline{2-3} 
\addlinespace
& Hanford & Livingston\\
\addlinespace
\hline
\hline
\addlinespace
Flat &  $10,685.11$ & $12,289.90$\\
\addlinespace
Hierarchical with background \\
follow-up & $712.89$ & $790.71$\\
\addlinespace
Hierarchical with only\\
foreground follow-up & $452.96$ & $529.21$\\
\addlinespace
\hline
\hline
\end{tabular}
\label{table:cpucorehour}
\end{table}

We also compared the computational efficiency of our search pipeline using a commonly employed metric in high-performance computing environments, that is, CPU core hours. We effectively distributed matched filtering operations across multiple CPU cores by leveraging the parallelization capabilities of the PyCBC toolkit. We found that the CPU core hour requirements for hierarchical search were consistently 14–15 times lower than those for flat search for both detectors, as shown in Table~\ref{table:cpucorehour}. This indicates a considerable reduction in computational resources and processing time associated with hierarchical searches. %

\section{\label{sec:conclusion}Conclusion and discussion}
Since its inception, the hierarchical search strategy has aimed to improve the efficiency of matched filtering by performing multi-stage searches using multiple banks of varying densities. Although it offers several advantages, especially in conducting computationally expensive searches for compact binaries such as sub-solar masses, accurately estimating an unbiased background has proven to be a persistent challenge. In particular, there is a constant tug-of-war between optimizing the computational cost of matched filtering and obtaining a proper noise background for assigning significance to detected candidates.%

In this work, we have successfully addressed this challenge by introducing a minor computational tradeoff. In Sec.~\ref{sec:far_hierarchical}, we proposed a scheme to estimate an unbiased noise background using a hybrid approach that combines the time-shifted background distribution and empirically models its tail based on the assumption that the distribution contains only noise coincidences. The effectiveness of our scheme was thoroughly demonstrated in Sec.~\ref{sec:application}, where we provided compelling evidence that our estimated FARs of true events from O3 closely align with those obtained through flat search. This result indicates the reliability and accuracy of our method for capturing the true significance of GW events. Because the tail of the distribution was obtained by extrapolation, one may argue that the background estimate in this region may contain a certain level of inaccuracy. However, the tail corresponds to exceptionally robust detections and any minor discrepancies in the FAR for these occurrences are likely to hold little significance when making inferences about astrophysical implications. Note that the background for the lower-ranking statistic ($\lesssim 8.5$), where distinguishing signals from noise presents a challenge, our estimations are derived directly from the time-slides, without involving any extrapolation, and hence, should be as reliable as the flat search. If necessary, this lower-ranking statistic region can be expanded further, perhaps up to a ranking statistic of approximately $\sim 9$. This expansion would involve increasing the number of time shifts, incurring additional computation time. However, despite this increase in computational demand, the overall computational efficiency of the entire search is still significantly advantageous. %

The significance of employing hierarchical search strategies for low-chirp mass sources becomes even more apparent based on our VT comparison results, as presented in Sec. ~\ref{sec:sensitivity_comparison}. We observed that with the current detection statistics and template bank configurations, the sensitivity of detecting lower-mass binaries is similar to that of flat search. However, in the case of higher chirp mass ranges, the sensitivity is reduced owing to the sparsity of templates in both the coarse bank and nbhd bank. Consequently, precise measurements of masses and spins are challenging for very short-duration signals with a hierarchical search. Thus, it is advisable to consider a more targeted search approach for such parameter ranges. Because the total number of templates in higher masses is much smaller than that for lower masses, we recommend performing the standard flat search for high masses, which adds an insignificant amount to the total computation cost. Incidentally, dedicated searches focused on high-mass binaries are performed routinely~\cite{gwtc3_paper}. Therefore, no additional effort may be necessary to improve computational efficiency in the above context for high-mass binaries. %

Although our approach has a minor reduction in computational gains compared to the previous implementation~\cite{kanchan_hierarchical}, where the background is estimated by scaling, this tradeoff is more than justified. Our present pipeline is robust, yet it substantially reduces the computational cost of matched filtering by an order of magnitude, and we find a computational speed-up of $\sim 13$ for O3 data. This ensures that the proposed method is efficient and practical for real-world GW detection.%

It is important to note that although our proposed method is effective for cases where multiple detectors detect GW signals, it may require careful attention and consideration for situations where the signal is captured by only one detector. In such cases, our background estimation may be susceptible to non-Gaussian artifacts caused by signal contamination. To address this potential issue, it is crucial to carefully extract and account for contamination, thus ensuring the robustness and reliability of our method, even in such challenging scenarios.%

In conclusion, our study introduces a robust approach for obtaining an unbiased background in a hierarchical search, which is a powerful and efficient tool for GW search. By effectively balancing the computational efficiency and accurate background estimation, our method paves the way for new avenues of research in the study of compact binaries.

\begin{acknowledgments}
The authors would like to express their sincere gratitude for the generous provision of computational resources by the IUCAA LDG cluster Sarathi and the support received from the LIGO Laboratory and National Science Foundation Grants. Additionally, the authors would like to thank Bhooshan Gadre for his invaluable insights into the work. K. S. also thanks Deepali Agarwal for useful discussions. This material is based upon work supported by NSF's LIGO Laboratory which is a major facility fully funded by the National Science Foundation. The hierarchical search pipeline employed PyCBC version 1.16.13, built upon the foundational frameworks of LALSuite~\cite{lalsuite_ref}, NumPy~\cite{numpy_ref}, SciPy~\cite{scipy_ref}, and Astropy~\cite{astropy_ref}. Furthermore, K. S. wishes to acknowledge the Inter-University Centre of Astronomy and Astrophysics (IUCAA), India, for the fellowship support. S. M. acknowledges the Department of Science and Technology (DST), Ministry of Science and Technology, India, for the support provided under the esteemed Swarna Jayanti Fellowships scheme. Lastly, S. V. D. expresses gratitude for the support received from the Senior Scientist Platinum Jubilee Fellowship, conferred by the National Academy of Sciences, India (NASI). This manuscript has been assigned a LIGO Document No. LIGO-P2300270.

\end{acknowledgments}

\appendix

\section{\label{apex:motivation} Analytical calculation for coincident noise distribution in Gaussian noise}
In this section, we attempt to understand the coincident noise distribution in the case of stationary Gaussian noise. The analytical calculation presented in this section substantiates the anticipated presence of linearly falling background distribution that we discussed in Sec.~\ref{sec:far_hierarchical}. This exercise aims to gain insights into the statistical distribution of detection statistics, particularly the quadrature sum of matched filter SNRs, in the context of Gaussian-colored noise. %

Suppose we aim to detect a GW signal embedded in stationary Gaussian noise, characterized by $N$ independent random variables. To accomplish this, we perform matched filtering on a data segment with a length of 512 s, sampled at 2048 Hz, using a template that consists of two orthonormal polarizations, $h_{o}$ and $h_{\pi/2}$. For the sake of simplicity, we assume that the occurrence of these signals is extremely rare within the data. The output of the matched filter can be regarded as the projection of the data vector onto the templates, as given by:
\begin{equation}
\rho = \sqrt{c_{o}^2 + c_{\pi/2}^2},
\end{equation}
where $\rho$ represents the SNR obtained from the matched filter, while $c_{o}$ and $c_{\pi/2}$ denote the projection coefficients onto $h_{o}$ and $h_{\pi/2}$, respectively. Since the two polarizations are orthonormal, $c_{o}$ and $c_{\pi/2}$ are normally distributed. The matched filter generates $N'$ samples of $\rho$ over time, and these samples follow a Rayleigh distribution.%

The samples produced by the matched filtering process are typically weakly correlated. Due to the weak correlations, we can assume that the cyclic operation of matched filtering generates $N < N'$ statistically independent Rayleigh variables. In such cases, the probability distribution of the maximum of $N$ independent Rayleigh variables is given by:
\begin{equation}\label{eq:pn}
    p_{N}(\rho) = N~\rho~e^{-\rho^2/2}~(1-~e^{-\rho^2/2})^{N-1}.
\end{equation}
The above pdf can be easily obtained by taking the product of $N$ Rayleigh distribution functions and then differentiating the product. For a large value of $N$, Eq.~(\ref{eq:pn}) can be approximated to a simpler form amenable to easy analytical manipulations as
\begin{equation}\label{eq:pdf_approx}
    p(\rho)\simeq N ~\rho~e^{-(\rho^2/2~+~ N ~e^{-\rho^2/2})} \,.
\end{equation}

\begin{figure}[ht!]
    \centering
    \includegraphics[scale=0.24]{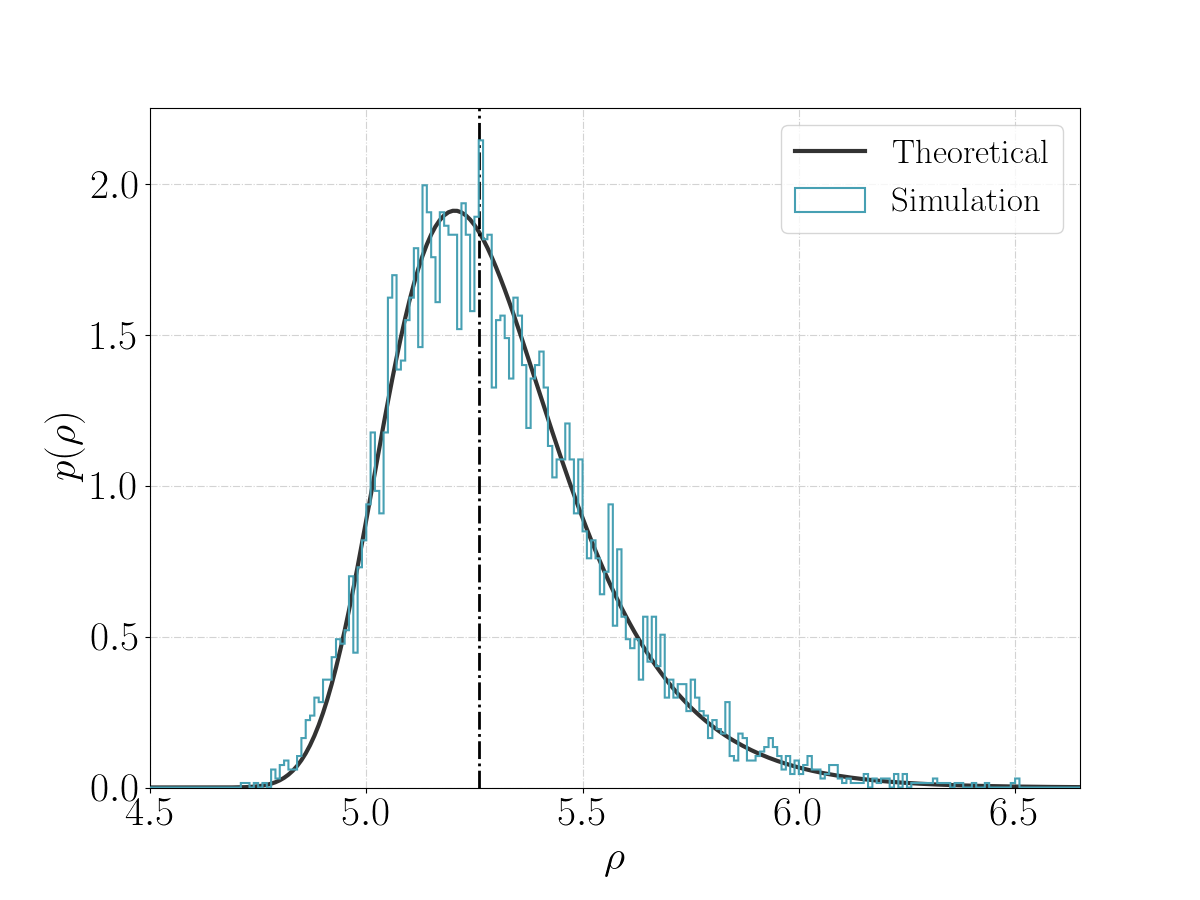}
    \caption{Probability density function of $\rho$ represented as a histogram (in blue). An ideal Rayleigh distribution is overlaid in black, providing a visual comparison. The vertical dotted line represents $\rho=5.3$.}
    \label{fig:snr_rayleigh}
\end{figure}

Let $y_1$ and $y_2$ represent two Rayleigh deviates corresponding to the SNRs from Hanford and Livingston, respectively. The probability distribution functions for $y_1$ and $y_2$ based on Eq. (\ref{eq:pdf_approx}) are given by,

\begin{align}\label{eq:1}
    p(y_1)= N_{1}~y_1~e^{-(y_1^2/2~+~ N_{1}  e^{-y_1^2/2})} \\
    p(y_2)= N_{2}~y_2~e^{-(y_2^2/2~+~ N_{2}  e^{-y_2^2/2})},
\end{align}
where $N_1$ and $N_2$ represent the maximum number of independent Rayleigh variables in each detector.  

Let us define the network SNR to be the coincident statistic in this case. Therefore, if we set $N = N_1 = N_2$, then the joint probability distribution function, denoted as $g(y_1,y_2)$, can be expressed as the product of the individual PDFs for $y_1$ and $y_2$,
\begin{align}
     g(y_1,y_2)= p(y_1)~p(y_2) \\
     = \frac{N^2}{2}~y_1~y_2~e^{-\big\{\frac{1}{2}(y^2_1+y^2_2)~+N(e^{-y^2_1/2}+e^{-y^2_2/2}) \big\}}.
\end{align}

Consider solving the above problem in polar coordinates. That means if $y_1=\rho_c\cos{\phi}$ and $y_2=\rho_c\sin{\phi}$, $g(y_1,y_2)$ (i.e., $\rho_c = \sqrt{y_1^2 + y_2^2}$) will become,

\begin{align}\label{eq:pdf_grhophi}
g(\rho_{c},\phi) &= \frac{N^2}{2}\rho_{c}^2~\sin{2\phi}~e^{-\rho_{c}^2/2} \nonumber \quad \quad \\
& \quad \times \bigg[e^{-N\big\{e^{-\frac{\rho_{c}^2}{4}(1+\cos{2\phi})}+e^{-\frac{\rho_{c}^2}{4}(1-\cos{2\phi})} \big\}}\bigg].
\end{align}

To obtain the joint PDF as a function of $\rho_{c}$, we marginalize Eq.~(\ref{eq:pdf_grhophi}) over $\phi$, that means,

\begin{align}
    g(\rho_{c}) = \frac{N^2}{2}\rho_{c}^3~e^{-\rho_{c}^2/2} \nonumber \quad \\
     \times \int^{\pi/2}_{0} d\phi~ \sin{2\phi}~\bigg[e^{-N_{c}~\big\{e^{-\frac{\rho_{c}^2}{4}(1+\cos{2\phi})}+e^{-\frac{\rho_{c}^2}{4}(1-\cos{2\phi})} \big\}}\bigg].
\end{align}

Setting $\cos{2\phi}=u$,
\begin{align}
    g(\rho_c) = \frac{N^2}{4}\rho_{c}^3~e^{-\rho_{c}^2/2} \int^{1}_{-1} du~e^{-N~\big\{e^{-\frac{\rho_{c}^2}{4}(1+u)}+e^{-\frac{\rho_{c}^2}{4}(1-u)} \big\}} \nonumber\\
    =  \frac{N^2}{2}\rho_{c}^3~e^{-\rho_{c}^2/2} \int^{1}_{0} du~e^{-N~\big\{e^{-\frac{\rho_{c}^2}{4}(1+u)}+e^{-\frac{\rho_{c}^2}{4}(1-u)} \big\}}.  
    \label{eq:inte}
\end{align}

\begin{figure}[ht!]
    \centering
    \includegraphics[scale=.25]{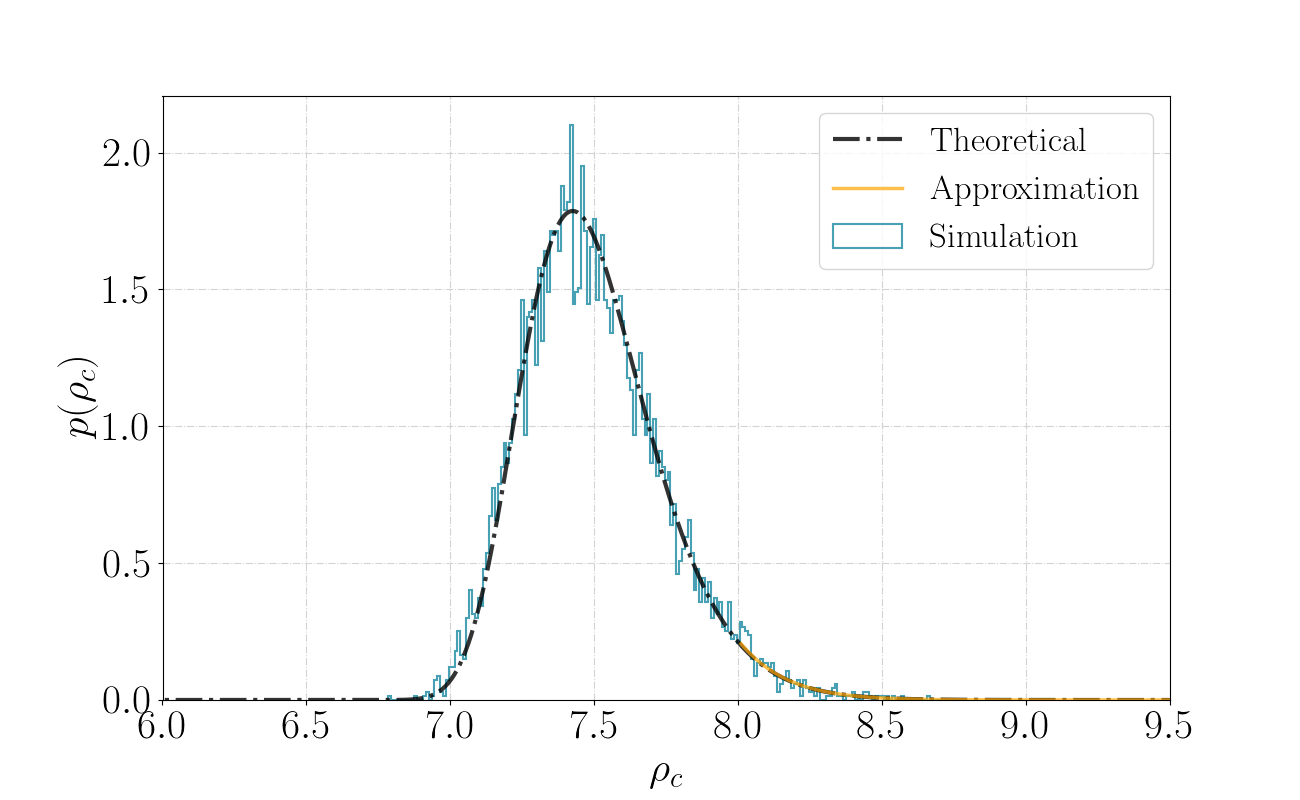}
    \caption{Probability density function for network SNR as a histogram obtained from simulations (blue) and by numerically integrating Eq.~\ref{eq:inte} (black).}
    \label{fig:plot_pdf_gaussian}
\end{figure}

It can be shown that the peak of this distribution occurs at $\sim \sqrt{4 \log N}$. Although the above integration can be numerically performed, we derive a semi-analytical fit for the tail of the distribution at large values of $\rho_{c} \gg \sqrt{4\log~N}$, where it is observed that the term $f(u) = e^{-N\frac{\rho_{c}^2}{4}(1-u)}$ dominates the integral. The function $f(u)$ is approximately shaped like a trapezium: for low values of $u$, $f(u) \simeq 1$ and $\longrightarrow 0$ as $u \longrightarrow 1$. We write the integral as $J$:
\begin{equation}
    J = \int^{1}_{0}du~f(u) \,.
\end{equation}
We approximate $f(u)$ using the trapezoidal rule. In this approximation, we set $f(u_o) = \frac{1}{2}$ at the mid-height of the trapezium, which has a unit height. Solving for $u_o$ yields the integral $J$,
\begin{equation}
    u_o = 1 - \frac{4}{\rho^2_c} (\log~N - \log~\log2) \equiv J \,.\label{eq:approximation_inte}
\end{equation} 

\begin{figure}[ht!]
    \centering
    \includegraphics[scale=.30]{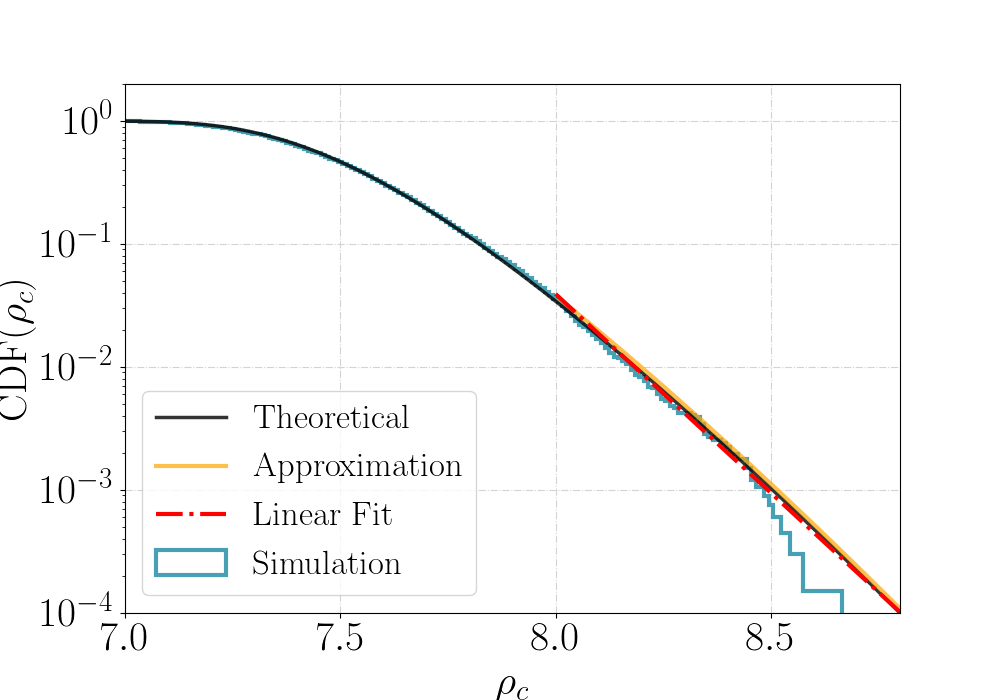}
    \caption{Cumulative density distributions obtained by integrating Eqs.\ref{eq:inte} (black) and~\ref{eq:approximation_inte} (orange), along with a histogram (blue) from simulated results. A linear fit (red) is overlaid in the logarithmic scale for comparison.}
    \label{fig:plot_cdf_gaussian}
\end{figure}

From Figure~\ref{fig:plot_pdf_gaussian}, it is evident that the approximation derived from Eq.~\ref{eq:approximation_inte} aligns well with Eq.~\ref{eq:inte} and the network SNR distribution obtained from the simulation. This observation prompts us to examine the cumulative distribution, as depicted in Fig.~\ref{fig:plot_cdf_gaussian}.%

Notably, the tail of the cumulative density function displayed in Fig.~\ref{fig:plot_cdf_gaussian}, derived from Eqs.~\ref{eq:inte} and~\ref{eq:approximation_inte}, exhibits a characteristic pattern that can be approximated by linearly decreasing values of ranking statistics when plotted on a logarithmic scale. This finding forms a benchmark for effectively modeling background distribution acquired through time-shifting triggers in real analysis. 

This exercise shows that the tail of the distribution in the general case has similar features as in this tractable case of Gaussian noise.

\nocite{*}


\bibliography{apssamp}

\end{document}